\documentclass[preprint,12pt]{article}
\usepackage{amsfonts}
\usepackage{amsmath}
\usepackage{amssymb}

\begin{document}
\title{Non-linear Liouville and Shr\"{o}dinger equations in phase-space}
\author{M.C.B. Fernandes$^{1}$, F.C. Khanna$^{2,3}$, M.G.R Martins$^{4}$, \\
A.E. Santana$^{1,2}$, J.D.M. Vianna$^{1,4}$ \\
$^{1}$Instituto de F\'{\i}sica, Universidade de Bras\'{\i}lia,\\
70910-900, Bras\'{\i}lia, DF, Brazil. \\
$^{2}$Department of Physics, University of Alberta, \\
Edmonton, Alberta, T6C 4G9, Canada\\
$^{3}$TRIUMF, Vancouver, British Columbia V6T 2A3, Canada\\
$^{4}$ Instituto de F\'{\i}sica, Universidade Federal da Bahia,\\
40210-340, Salvador, Bahia, Brazil.}
\maketitle

\begin{abstract}

Unitary representations of the Galilei group are studied in phase
space, in order to describe classical and quantum systems.
Conditions to write in general form the generator of time
translation and Lagrangians in phase space are then established.
In the classical case, Galilean invariance provides conditions for
writing the Liouville operator and Lagrangian for non-linear
systems. We analyze, as an example, a generalized kinetic equation
where the collision term is local and non-linear. The quantum
counter-part  of such unitary representations are developed by
using the Moyal (or star) product. Then a non-linear
Schr\"{o}dinger equation in phase space is derived and analyzed.
In this case, an association with the Wigner formalism is
established, which provides a physical interpretation for the
formalism.
\end{abstract}

\section{Introduction}

 Phase space ($\Gamma$) is the natural manifold for formulation of the kinetic
 theory; and as such, is the basic starting point for exploring
symmetry. For classical particles and fields,  the Poisson
bracket, a symplectic two-forms in $\Gamma$,  is mapped in the Lie
product of Lie algebras~\cite{dir1,muk1}, giving rise to
representations of kinematical groups. In particular, unitary
representations constructed from a Hilbert space in $\Gamma$ are
of interest for their practical appeal: taking advantage of the
notion of linear space  to develop, for instance, perturbative
techniques.

For the Galilei group, describing non-relativistic systems,
unitary representations for classical statistical systems were
proposed by Sch\"{o}nberg~\cite{schem1,schem2,schem3}, who
introduced the notion of Fock-space in $\Gamma$ and the study of
unitary symplectic representations. Numerous other developments
then followed, to consider the Brownian motion, stochastic
processes, classical kinetic theory and generalizations of these
ideas to the quantum domain~\cite{Loinger}-\cite{Santana1}.

The phase space for quantum systems is introduced by the Wigner
function~\cite{wig1}-\cite{ViannaCezar}. In such an
approach  each operator, $A$, defined in the usual Hilbert space, $%
\mathcal{H}$, is associated with a function, $a_{W}(q,p)$, in
$\Gamma $~\cite{wig2}-\cite{moy2}. Then there is a mapping $\Omega
_{W}:A\rightarrow a_{W}(q,p)$, such that, the associative algebra
of operators defined in $\mathcal{H}$ turns out to be an
associative algebra in $\Gamma ,$ given by $\Omega
_{W}:AB\rightarrow a_{W}\ast b_{W},$ where the star (or
Moyal)-product, $\ast \,$, is given as
\begin{equation}
a_{W}\ast b_{W}=a_{W}(q,p)\exp \left[ \frac{i\hbar}{2}(\frac{\overleftarrow{%
\partial }}{\partial q}\frac{\overrightarrow{\partial }}{\partial p}-\frac{%
\overleftarrow{\partial }}{\partial p}\frac{\overrightarrow{\partial }}{%
\partial q})\right] b_{W}(q,p),     \label{dessa111}
\end{equation}%
This provides a non-commutative algebraic structure in the phase
space, that has been explored in different
ways~\cite{wig2}-\cite{Bohm2}. The study of unitary
representations of Lie groups in phase space for quantum systems
has been achieved~\cite{seb1}-\cite{seb22}, by using  the Weyl
operators, $\widehat{a}=a_{W}\ast $, that are introduced as a
mapping on functions $b_{W},$ such that
$\widehat{a}(b_{W})=a_{W}\ast b_{W}.$
 This
symplectic representation  provides a way to consider a
perturbative approach for Wigner functions based on symmetry
groups. One example is the $\lambda \phi ^{4}$ field theory in
phase-space, giving rise to a relativisitic kinetic equation with
a local Boltzmann-like collision term. It is important to
emphasize that, although associated with the Wigner formalism, the
symplectic representations have a Hamiltonian, and not a
Liouville, operator as generator of time translations.

From a conceptual standpoint, formulations of physics in phase
space are such that the generator of time translation, as the
classical and  quantum Liouvillian operators, is usually defined
by using the Hamiltonian. In both cases, it is necessary to know
the Hamiltonian first and then it is possible to proceed to the
Liouvillian formulation. This has been recognized as a hindrance
to exploring a variety of phenomena in kinetic theory and
stochastic problems, involving non-linear elements and
irreversibility~\cite{Misra,doi3,doi8,seb11,Bohm2}

By using unitary symplectic representations, we show here that
such a path to find the generator of time translation in phase
space is not necessary. With the   Galilean symmetries applied to
time evolution of physical states, we find algebraic relations,
which the generator of time translation must satisfy. Then it is
possible to infer the form of the classical Liouville operator
without previous knowledge of the Hamiltonian. The Liouville
operator is interpreted independently of the Hamiltonian form,
having a life of its own. In a similar way, the Hamiltonian in
phase space describing quantum systems is constructed. In
addition, by using the Hilbert space defined in $\Gamma$ and the
Galilei symmetry, we analyze the Lagrangian formalism. This
procedure opens numerous
 possibilities to introduce interactions and non-linear
effects in the kinetic theory.

We explore these possibilities by studying  a non-linear
Schr\"{o}dinger (or a Gross-Pitaevskii-like) equation in phase
space~\cite{GP1,GP2}. The association of this formalism with the
Wigner function is then discussed. In the case of classical
systems, we analyze a Liouville-like equation with a non-linear
source term. These non-linear equations are solved perturbatively,
showing a systematic procedure to use
 the group theory analysis to improve and to explore the kinetic
theory.

The paper is organized as follows. In Section 2, we briefly review
sympletic manifolds in order to define unitary representations in
phase space. In Section 3, we consider classical systems and the
non-linear Liouville equation. In Section 4, we study quantum
representations using the Moyal product. In Section 5, the
non-linear Schr\"{o}dinger equation in phase space is studied.
Finally, in Section 6, some concluding remarks are presented.

\section{Symplectic manifolds and Hilbert space}

Consider an analytical manifold $\mathbb{M}$ where each point is specified
by Euclidian coordinates $q^{i},$ with $i=1,2,3$. The coordinates of each
point in the cotangent-bundle $\Gamma =T^{\ast }\mathbb{M}$ is denoted by $%
(q^{i},p^{i})$. The space $\Gamma $ is equipped with a symplectic structure
by the 2-form
\begin{equation}
\omega =\sum\limits_{i=1}^{3}dq^{i}\wedge dp^{i}  \label{simp1}
\end{equation}%
Let us define the following,
\begin{equation}
\Lambda =\sum\limits_{i=1}^{3}\frac{\overleftarrow{\partial }}{\partial q^{i}%
}\frac{\overrightarrow{\partial }}{\partial p^{i}}-\frac{\overleftarrow{%
\partial }}{\partial p^{i}}\frac{\overrightarrow{\partial }}{\partial q^{i}},
\label{fasenova2}
\end{equation}%
such that for $C^{\infty }$ functions, $f(q,p)$ and $g(q,p),$ we have
\begin{equation}
\omega (f\Lambda ,g\Lambda )=f\Lambda g=\{f,g\},  \label{fasenova3}
\end{equation}%
where $\{f,g\}=\sum\limits_{i=1}^{3}(\frac{\partial f}{\partial q^{i}}\frac{%
\partial g}{\partial p^{i}}-\frac{\partial f}{\partial p^{i}}\frac{\partial g%
}{\partial q^{i}})$ is the Poisson bracket. We identify  vector fields in $%
\Gamma $ by
\begin{equation}
f\Lambda =X_{f}=\sum\limits_{i=1}^{3}(\frac{\partial f}{\partial q^{i}}\frac{%
\partial }{\partial p^{i}}-\frac{\partial f}{\partial p^{i}}\frac{\partial }{%
\partial q^{i}}),  \label{fasenova4}
\end{equation}%
where $  f=f(q^{i},p^{i}) \in C^{\infty }(\Gamma )$. The space
$\Gamma $, endowed with this symplectic structure, is called the
phase space.

In order to construct a Hilbert space over $\Gamma $, let $\mu $ be an invariant measure on the cotangent bundle. If $%
\varphi $ is a mapping: $\Gamma \rightarrow \mathbb{R}$ which is
measurable. Then we define the integral of $\varphi $ with respect
to $\mu $ as
\begin{equation}
\int_{\Omega }\varphi (\mathbf{z})d\mu (\mathbf{z}),
\end{equation}%
 where $\mathbf{z}\in \Gamma $. Let $%
\mathcal{H}(\Gamma )$ be a linear subspace of the space of $\mu
$-measurable functions $\psi :\Gamma \rightarrow \mathbb{C}$ which
are square integrable, such that
\begin{equation}
\int_{\Gamma }\mid \psi (\mathbf{z})\mid ^{2}d\mu (\mathbf{z})<\infty .
\end{equation}%
We equip $\mathcal{H}(\Gamma )$ with an inner product, $\langle
\cdot |\cdot \rangle $ by
\begin{equation}
\langle \psi _{1}|\psi _{2}\rangle =\int_{\Gamma }\psi
_{1}(q,p)^{\dagger }\psi _{2}(q,p)d\mu (q,p),
\end{equation}%
where we take $\mathbf{z}=(q^{i},p^{i})=(q,p)$, and $\psi (q,p)$ $\in $ $%
C^{\infty}(\Gamma) $ is such that
\begin{equation}
\int d^{3}pd^{3}q\psi ^{\dagger }(q,p)\psi (q,p)<\infty .
\end{equation}%
Then $\mathcal{H}(\Gamma )$  is a Hilbert space.

In this case, we have $\psi (q,p)=\langle q,p|\psi \rangle $, with
\begin{equation}
\int d^{3}pd^{3}q|q,p\rangle \langle q,p|=1,
\end{equation}%
such that, the kets $|q,p\rangle $ are defined from the set of operators $%
\bar{Q}$ and $\bar{P}$, such that
\begin{equation*}
\bar{Q}|q,p\rangle =q|q,p\rangle ,\ \ \ \ \bar{P}|q,p\rangle =p|q,p\rangle ,
\end{equation*}%
satisfying the commutation condition $[\bar{Q},\bar{P}]=0$. The
state of a system is described by functions $\phi (q,p)$, with the
condition
\begin{equation}
\langle \psi |\phi \rangle =\int d^{3}pd^{3}q\psi ^{\dagger }(q,p)\phi
(q,p)<\infty
\end{equation}%
This Hilbert space, $\mathcal{H}(\Gamma ),$ is taken as the
representation space to provide a general scheme to study unitary
representations of the
Galilei group. A unitary transformation in $\mathcal{H}(\Gamma )$ is the mapping $U:%
\mathcal{H}(\Gamma )\rightarrow \mathcal{H}(\Gamma )$ such that
$\langle \psi _{1}|\psi _{2}\rangle $ is invariant.

Observe that a general associative product in $\mathcal{H}(\Gamma
)$ is introduced  as a mapping $%
e^{ia\Lambda }=\ast :$ \ $\Gamma \times \Gamma \rightarrow \Gamma
,$ called the Moyal (or star) product, as given in
Eq.~(\ref{dessa111}), i.e.
\begin{equation}
f\ast g =f(q,p)e^{ia\Lambda } g(q,p),  \label{A98}
\end{equation}%
where $f$ and $g$ are functions in phase-space and $\partial
_{z}=\partial /\partial z$ $(z=p,q).$ The constant $a$ is used at
this point to fix units, without any special meaning. The usual
associative product is obtained by taking $a=0.$ In addition, to
each function, say $f(q,p)$, we introduce operators in the form
${\widehat{f}}=f(q,p)\ast $. Such an operator will be used as the
generator of unitary transformations.

In the following sections these two types of representations are
analyzed
explicitly. We take into account the Lie algebra for the Galilei group, $%
\mathfrak{g}$, given by
\begin{equation}
\left.
\begin{array}{ll}
\,[\widehat{L}_{i},\widehat{L}_{j}]=i \epsilon _{ijk\ }\widehat{L}_{k}, & \,[\widehat{K}%
_{i},\widehat{P}_{j}]=i a_{0}\delta _{ij}, \\
\,[\widehat{L}_{i},\widehat{K}_{j}]=i \epsilon _{ijk}\widehat{K}_{k}, & \,[\widehat{K}%
_{j},\widehat{H}]=i  \widehat{P}_{j}, \\
\,[\widehat{L}_{i},\widehat{P}_{j}]=i  \epsilon _{ijk}\widehat{P}_{k}, & \,[\widehat{P}%
_{i},\widehat{P}_{j}]=0, \\
\,[\widehat{L}_{i},\widehat{H}]=0, & \,[\widehat{P}_{i},\widehat{H}]=0, \\
\,[\widehat{K}_{i},\widehat{K}_{j}]=0, &
\end{array}%
\right\}  \label{12nov1}
\end{equation}%
where $\widehat{P},\widehat{K},\widehat{L}$and$\widehat{H}$ are
the generators of translations, boost, rotations and  time
translations, respectively. The constant $a_{0}$ is the central
extension of the group. Defining two operators $Q$ and $P$ that
are transformed by the boost according to
\begin{eqnarray}
\exp \left( -iv\cdot \widehat{K}\right) Q_{j}\exp \left( iv\widehat{K%
}\right) &=&Q_{j}+v_{j}t  \label{MAIS5} \\
\exp \left( -iv\cdot \widehat{K}\right) P_{j}\exp \left( iv\cdot
\widehat{K}\right) &=&P_{j}+mv_{j}, \label{mais66}
\end{eqnarray}%
then the physical content of the algebra become obvious. The
operators $Q$ and $P$ are interpreted as  position and momentum,
respectively. These are basic relations used to derive physical
conditions to study  representations. In the following sections we
consider classical and quantum representations.

\section{Symplectic classical mechanics}

Let us consider unitary representations in phase space describing
a classical system. This is achieved by using the vector field in
phase space given in Eq.~(\ref{fasenova4}), i.e. we introduce
unitary operators with the definition $\widehat{f}=iX_{f}$. We
consider the following set of operators:
\begin{eqnarray*}
{\widehat{P}}_{i} &=&iX_{p_{i}}=-i{\frac{\partial }{\partial q_{i}},} \\
{\widehat{K}}_{i} &=&iX_{K_{i}}=im{\frac{\partial }{\partial p_{i}}}-it{\frac{%
\partial }{\partial q_{i}}}, \\
{\widehat{J}}_{i} &=&iX_{J_{i}}={\widehat{L}}_{i}+{\widehat{S}}_{i}, \\
{\widehat{H}} &=&i{\frac{\partial }{\partial t}},
\end{eqnarray*}%
where%
\begin{equation*}
{\widehat{L}}_{i}=iX_{L_{i}}=i\varepsilon _{ijk}\left( q_{k}{\frac{\partial }{%
\partial q_{j}}}+p_{k}{\frac{\partial }{\partial p_{j}}}\right) ,
\end{equation*}%
with $L_{i}=\varepsilon _{ijk}q_{j}p_{k}.$ The boost operator is
constructed with the function $K_{i}=mq_{i}-tp_{i}.$ The operators
${\widehat{S}}_{i}$ are the spin operators (a representation of
$SO(3)$ such that $\widehat{S}$ commutes
with every operator defined on the phase space. We take here ${\widehat{S}}%
_{i}=0$. This set of operators fulfills the relations given in Eq.~(\ref%
{12nov1}) with $a_{0}=0$.

Let us define the linear operators $P_{i}$ and $Q_{i}$   by%
\begin{equation*}
P_{i}|p_{i},q_{i}\rangle =p_{i}|p_{i},q_{i}\rangle ,\ \ \
Q_{i}|p_{i},q_{i}\rangle =q_{i}|p_{i},q_{i}\rangle ,
\end{equation*}%
such that $\langle q,p|\theta \rangle =\theta (p,q)$ is a vector
in the phase space representation, which is a vector in
$\mathcal{H}(\Gamma )$. Notice that $%
[P_{i},Q_{j}]=0.$  Let us evaluate the physical consequences of
this representation.

The operators $P$ and $Q$ are interpreted as the momentum and
position operators, since they satisfy the Galilei boost
conditions, namely
\begin{equation}
\langle \theta |\exp {(-iv{\widehat{K}})}Q\exp
{(iv{\widehat{K}})}|\phi \rangle =\langle \theta |Q|\phi \rangle
+vt\langle \theta |\phi \rangle , \label{boost2}
\end{equation}%
and
\begin{equation}
\langle \theta |\exp {(-iv{\widehat{K}})}P\exp
{(iv{\widehat{K}})}|\phi \rangle =\langle \theta |P|\phi \rangle
+mv\langle \theta |\phi \rangle . \label{boost3}
\end{equation}%
where $|\theta \rangle $ and $|\phi \rangle $ (${\in }\
\mathcal{H}$) are arbitrary states of the system, and
$[{\widehat{P}}_{i},Q_{j}]=-i\delta _{ij}$. Then, $L$ is the
angular momentum, and $H$ is the Hamiltonian.

Since $[{\widehat{P}}_{i},{\widehat{K}}_{j}]=0$, then
$a_{0}=0 $ in Eq.~(\ref{12nov1}).  However, by introducing the c-number
operator $K=mQ-tP,$ we have $[%
{\widehat{P}}_{i},K_{j}]=[P_{i},{\widehat{K}}_{j}]=-im\delta
_{ij}$. Taking this relation, together with Eqs.~(\ref{boost2})
and (\ref{boost3}), we find that the constant $m$ is mass. These
relations among $K$, $P$ and $Q$ are similar to those used in
quantum mechanics, but  $Q$ and $P$ commute with one another,
since they describe the position and momentum of a classical
system.

The expectation value of a dynamical variable $\bar{A}$ in a state $|\theta
\rangle $ is defined by
\begin{equation}
\langle \bar{A}\rangle =\langle \theta |\bar{A}|\theta \rangle .
\label{average}
\end{equation}%
On the other hand, the temporal evolution of $\bar{A}$ is given by
\begin{equation}
\langle \theta _{0}|\exp {(it{\widehat{H}})}{\bar{A}}\exp ({(-it{\widehat{H}})}%
|\theta _{0}\rangle =\langle \theta _{0}|{\bar{A}}(t)|\theta
_{0}\rangle , \label{heisenberg1}
\end{equation}%
where $\widehat{H}$ is called the Liouvillian. Therefore, we have
defined a Heisenberg picture for the temporal evolution of the
dynamical variables, and from Eq.~(\ref{heisenberg1}) we obtain
\begin{equation}
i{\partial _{t}}{\bar{A}}=[\bar{A},\widehat{H}].
\label{heisenberg2}
\end{equation}

In the Schr\"{o}dinger picture, the evolution of the state is
given by
\begin{equation}
i{\partial _{t}|{\theta }(t)}\rangle =\widehat{H}|\theta
(t)\rangle , \label{liouvilleeq}
\end{equation}%
Using the orthonormality of the states $|q,p\rangle$,
\begin{equation*}
\langle q,p|q^{\prime },p^{\prime }\rangle =\delta (q-q^{\prime })\delta
(p-p^{\prime })\,\,\text{and}\,\,\int |q,p\rangle \langle q,p|dqdp=1,
\end{equation*}%
we write
\begin{equation}
i\partial _{t}\theta (q,p;t)\ =\int \langle
q,p|\widehat{H}|q^{\prime },p^{\prime }\rangle \,\langle q^{\prime
},p^{\prime }|\theta (t)\rangle dq^{\prime }dp^{\prime },
\label{car253}
\end{equation}%
where $|\theta (t)\rangle $ is in $\mathcal{H}(\Gamma )$ and $\theta
(q,p;t)=\langle q,p|\theta (t)\rangle $. Assuming
\begin{equation*}
\langle q,p|\widehat{H}|q^{\prime },p^{\prime }\rangle =\delta
(q-q^{\prime })\delta (p-p^{\prime })\langle
q,p|\widehat{H}|q,p\rangle ,
\end{equation*}%
we have%
\begin{equation}
i\partial _{t}\theta (q,p;t)\ =\mathcal{L}(q,p)\theta (q,p;t),
\label{liouaug08}
\end{equation}%
where $\mathcal{L}(q,p)=\langle q,p|\widehat{H}|q,p\rangle $ is
the classical Liouville operator, and the connection with the
Liouville equation is to be derived. A simple solution is to
consider $\mathcal{L}(q,p)=iX_{H}=i\{H,.\}$ , where
$H=p^{2}/2m$~\cite{Arthur}. A central point here is to find a
general form for $\mathcal{L}(q,p).$ We solve this problem by
using the symmetry properties of the Galilei group. To proceed
further, let us discuss some additional aspects about this
symplectic representation for classical physics, as it was first
proposed by Sch\"{o}nberg \cite{schem1,schem2,schem3}. A set of
rules for physical interpretation rules has to be established. For
an $n$-particle system described in the phase space, these rules
are the following.
\begin{itemize}
\item[(i)] The states of an $n$-particle system are vectors in Hilbert state $%
H(\Gamma ).$   Each vector is given by a wave function $\theta
_{n}=\theta (z_{1},\cdot \cdot \cdot ,z_{n})$ with
$(z_{i}:=(q_{i},p_{i}))$, such that, the probability density  in
the classical phase space is written
as $f_{n}:=f(z_{1},\cdot \cdot \cdot ,z_{n})=|\theta _{n}|^{2}$. The state $%
\theta _{n}$ satisfies Eq. (\ref{liouaug08}), that is written in
the notation of $n$-particle systems, explicitly, as
\begin{equation}
\frac{\partial \theta _{n}}{\partial t}=\{H,\theta _{n}\}_{n}=-i\mathcal{L}%
_{n}\theta _{n}  \label{le}
\end{equation}%
where,
\begin{equation}
\{H,\theta _{n}\}_{n}=\sum_{i=1}^{3n}(\frac{\partial H_{n}}{\partial q_{i}}%
\frac{\partial \theta _{n}}{\partial p_{i}}-\frac{\partial H_{n}}{\partial
p_{i}}\frac{\partial \theta _{n}}{\partial q_{i}})\   \label{hvf}
\end{equation}%
and $H_{n}$ is the Hamiltonian of a classical $n$-particle system.

\item[(ii)] To each physical quantity $a(z_{1},\cdot \cdot \cdot ,z_{n})\equiv a(q,p)$ in
phase space, two hermitian operators on the space $%
\mathcal{H}(\Gamma )$ are associated; i.e. a diagonal operator $A$
and a differential operator
$\widehat{A}=i\sum_{i=1}^{3n}(\frac{\partial a}{\partial
p_{i}}\frac{\partial }{\partial q_{i}}-\frac{\partial a}{\partial q_{i}}%
\frac{\partial }{\partial p_{i}})\equiv i\{a,\cdot \}$. The
operators $A$ are the usual physical observables, whereas
operators of type $\widehat{A}$ are called dynamical generators of
symmetries. The possible values of a physical quantity represented
by $A$ are its eigenvalues. It follows that the average value of
the quantity $A$ in the state $\theta _{n}$ is $\langle A\rangle
=\int dz\theta _{n}^{\ast }A\theta _{n}=\int dz|\theta
_{n}|^{2}a(q,p)$; i.e, the classical result.

\item[(iii)] In the symplectic representation, we introduce three pictures
for the state vector $|\theta (t)\rangle $ as well as for
operators of the theory. They are, the classical Schr\"{o}dinger
picture, the classical Heisenberg picture and the classical
interaction picture \cite{schem1,Arthur}. We have used above the
classical Heisenberg and Schr\"{o}dinger pictures, in the analysis
of the symplectic representation.
\end{itemize}

At this point it is instructive to compare this representation for
classical systems with the usual formulation of quantum physics.
In  quantum mechanics, symmetry transformations are represented by
unitary operators acting on the Hilbert space~\cite{Ballentine}.
The Hamiltonian operator, for example, represents the
infinitesimal generator of time translations whereas momentum
generates spatial translations, and so on. This can be formulated
in terms of  representations of the Galilei group in the Hilbert
space. In the classical symplectic representation, on the other
hand, symmetry operations are mapping on $\mathcal{H}(\Gamma ),$
and one example is the Liouville operator $\mathcal{L}_{n},$ that
is the generator of time translations. This conclusion is reached
by using for instance the Hamiltonian $H_{n}$. Using Galilean
invariance, however, we have to show the general algebraic
conditions for the representation of time translation generator,
$\mathcal{L}_{n},$ irrespective of any Hamiltonian formulation
previously assumed.

Let $|\theta _{n}\rangle \in \mathcal{H}(\Gamma )$ be an arbitrary $n$%
-particle state prepared by an observer $\mathcal{O}$ at the instant $t_{0}$%
, and let $|\theta _{n};\mathbf{v}\rangle $ be the state having
the same properties at time $t_{0}$ insofar as an observer
$\mathcal{O} ^{'}$, who is moving with velocity $\mathbf{v}$
relative to $\mathcal{O}$, is concerned. Let us assume that at
$t=0$ the two coordinate frames coincide. The
expectation value of the operators $\mathbf{Q}_{i}:=(Q_{ix},Q_{iy},Q_{iz})$%
, $\mathbf{P}_{i}:=(P_{ix},P_{iy},P_{iz})$ and $\widehat{F}$ (an
arbitrary physical quantity in phase space) in these states is
then related by the Galilei transformations,
\begin{eqnarray}
\langle \theta _{n};\mathbf{v}|\mathbf{Q}_{i}|\theta _{n};\mathbf{v}\rangle
&=&\langle \theta _{n}|\mathbf{Q}_{i}|\theta _{n}\rangle +\mathbf{v}t_{0}
\label{g1} \\
\langle \theta _{n};\mathbf{v}|\mathbf{P}_{i}|\theta _{n};\mathbf{v}\rangle
&=&\langle \theta _{n}|\mathbf{P}_{i}|\theta _{n}\rangle +m_{i}\mathbf{v}
\label{g2} \\
\langle \theta _{n};\mathbf{v}|\widehat{\mathbf{F}}_{i}|\theta _{n};\mathbf{v%
}\rangle &=&\langle \theta _{n}|\widehat{\mathbf{F}}_{i}|\theta _{n}\rangle
+\Delta _{F}\mathbf{v},  \label{g3}
\end{eqnarray}%
with  $\Delta _{F}$ being a quantity to be determined for each $%
\widehat{\mathbf{F}}$. Considering an infinitesimal Galilei transformation,
we replace $\mathbf{v}$ by $\delta \mathbf{v}$ and define the infinitesimal
unitary operator
\begin{equation}
\widetilde{\Gamma }(t_{0},\delta \mathbf{v})=1-i\delta \mathbf{v}\cdot
\widehat{\mathbf{K}},\;\; \widehat{\mathbf{K}}=\widehat{\mathbf{K}}%
^{\dagger },  \label{igt}
\end{equation}%
by requiring that $\widetilde{\Gamma }|\theta _{n}\rangle =|\theta
_{n};\delta \mathbf{v}\rangle $. We assume that $\widehat{\mathbf{K}}=\mathbf{K}(\mathbf{q}_{i},\mathbf{p}%
_{i},\frac{\partial }{\partial \mathbf{q}_{i}},\frac{\partial
}{\partial \mathbf{p}_{i}}) $, in Eq. (\ref{igt}).

It is to be noted that for $\widehat{F}=iX_{f}=i\{f,\cdot \}$ and $\widehat{A}%
=iX_{a}=i\{a,\cdot \},$ we have:
\begin{equation}
\lbrack A,\widehat{F}]_{-}=i\{a(q,p),f(q,p)\},  \label{cpb1}
\end{equation}%
\begin{equation}
\lbrack i\{f,\cdot \},i\{a,\cdot \}]=-\{\cdot ,\{a,f\}\}  \label{cpb2}
\end{equation}%
and
\begin{equation}
\lbrack A,F]=0,  \label{cpb3}
\end{equation}%
where $A$ and $F$ are diagonal operators in $\mathcal{H}(\Gamma
)$.

By virtue of Eqs.~(\ref{g1}),(\ref{g2}) and (\ref{g3}), and the
definition given in Eq. (\ref{igt}), we get
\begin{eqnarray}
t_{0}\delta \mathbf{v} &=&-i[\mathbf{Q}_{i},\delta \mathbf{v}\cdot \widehat{%
\mathbf{K}}],  \label{a1} \\
m_{i}\delta \mathbf{v} &=&-i[\mathbf{P}_{i},\delta \mathbf{v}\cdot \widehat{%
\mathbf{K}}],  \label{a2} \\
\Delta _{F}\delta \mathbf{v} &=&-i[\widehat{F},\delta
\mathbf{v}\cdot \widehat{\mathbf{K}}].  \label{a3}
\end{eqnarray}%
Non-trivial solutions of these equations are found by taking $\widehat{%
\mathbf{K}}=iX_{\mathbf{k}}=i\{\mathbf{k},\cdot \}$, where $\mathbf{k=k(q},%
\mathbf{p)}$ is a vector-function in phase space. Hence,
\begin{equation}
t_{0}\delta \mathbf{v}=-i[\mathbf{Q}_{i},i\{\mathbf{k},\cdot \}]\cdot \delta
\mathbf{v}=\frac{\partial }{\partial \mathbf{p}_{i}}\delta \mathbf{v}\cdot
\mathbf{k},  \label{nc1}
\end{equation}%
\begin{equation}
m_{i}\delta \mathbf{v}=-i[\mathbf{P}_{i},i\{\mathbf{k},\cdot \}]\cdot \delta
\mathbf{v}=-\frac{\partial }{\partial \mathbf{q}_{i}}\delta \mathbf{v}\cdot
\mathbf{k},  \label{nc2}
\end{equation}%
\begin{equation}
\Delta _{\mathbf{Q}_{i}}\delta \mathbf{v}=-i\delta \mathbf{v}\cdot \{\cdot ,%
\frac{\partial \mathbf{k}}{\partial \mathbf{p}_{i}}\},  \label{nc3}
\end{equation}%
\begin{equation}
\Delta _{\mathbf{P}_{i}}\delta \mathbf{v}=i\delta \mathbf{v}\cdot \{\cdot ,%
\frac{\partial \mathbf{k}}{\partial \mathbf{q}_{i}}\},  \label{nc4}
\end{equation}%
where we have used for $\widehat{F}$ the operators $\widehat{\mathbf{Q}}_{i}$
and $\widehat{\mathbf{P}}_{i}$.

As $\delta \mathbf{v}$ is arbitrary, these equations reduce to:%
\begin{eqnarray*}
t_{0} &=&\frac{\partial k_{x}}{\partial p_{i_{x}}}, \\
m_{i} &=&-\frac{\partial k_{x}}{\partial q_{i_{x}}}, \\
i\Delta _{x} &=&\frac{\partial ^{2}k_{x}}{\partial p_{i_{x}}^{2}}\frac{%
\partial }{\partial q_{i_{x}}}-\frac{\partial ^{2}k_{x}}{\partial
q_{i_{x}}\partial p_{i_{x}}}\frac{\partial }{\partial p_{i_{x}}}, \\
-i\Delta _{p_{x}} &=&\frac{\partial ^{2}k_{x}}{\partial p_{i_{x}}\partial
q_{i_{x}}}\frac{\partial }{\partial q_{i_{x}}}-\frac{\partial ^{2}k_{x}}{%
\partial q_{i_{x}}^{2}}\frac{\partial }{\partial p_{i_{x}}}.
\end{eqnarray*}
A solution for these equations is derived considering that the Poisson
brackets of coordinates $\mathbf{q}_{i}$ and momenta $\mathbf{p}_{i}$ are
invariant under Galilei transformations, i.e. $\Delta _{x}=\Delta _{p_{x}}=0$%
, such that,
\begin{equation}
\mathbf{k}_{t_{0}}=t_{0}\mathbf{p}-M\mathbf{R},\;\;\;\mathbf{R}=\frac{%
\sum_{i}m_{i}\mathbf{q}_{i}}{M},\;\;M=\sum_{i}m_{i},\;\;\mathbf{p}=\sum_{i}%
\mathbf{p}_{i}  \label{s1}
\end{equation}

Let $U(t_{1},t_{0})=e^{-i\mathcal{L}_{n}(t_{1}-t_{0})}$ be the
time-evolution operator on $\mathcal{H}(\Gamma )$, such that
\begin{equation}
U(t_{1},t_{0})\widetilde{\Gamma }(t_{0},\mathbf{v})=\widetilde{\Gamma }%
(t_{1},\mathbf{v})U(t_{1},t_{0}).  \label{te1}
\end{equation}%
This equation imposes a condition on the operator
$\mathcal{L}_{n}$. In order to obtain this condition it suffices
to replace $\mathbf{v}$ by the infinitesimal $\delta \mathbf{v}$.
Therefore, with the aid of Eq.~(\ref{igt}), we obtain
\begin{equation}
U(t_{1},t_{0})(1+\delta \mathbf{v}\cdot \{\mathbf{k}_{t_{0}},\cdot
\})=(1+\delta \mathbf{v}\cdot \{\mathbf{k}_{t_{1}},\cdot \})U(t_{1},t_{0}).
\label{ite}
\end{equation}%
Using Eq.~\ref{s1}), this equation reads
\begin{eqnarray}
-t_{0}U(t_{1},t_{0})\frac{\partial }{\partial \mathbf{q}_{i}}
&=&U(t_{1},t_{0})m_{i}\frac{\partial }{\partial \mathbf{p}_{i}}  \notag \\
&&-t_{1}\frac{\partial }{\partial \mathbf{q}_{i}}U(t_{1},t_{0})-m_{i}\frac{%
\partial }{\partial \mathbf{p}_{i}}U(t_{1},t_{0}).  \label{s2aug}
\end{eqnarray}%
Since the system is invariant under space translations, the operator $%
\partial /\partial \mathbf{q}_{i}$commutes with $U(t_{1},t_{0})$. Moreover,
with $\delta t\simeq 0$, from Eq. (\ref{s2aug}), we find
\begin{equation}
\lbrack m_{i}\frac{\partial }{\partial \mathbf{p}_{i}},1-i\mathcal{L}%
_{n}\delta t]=-\delta t(1-i\mathcal{L}_{n}\delta t)\frac{\partial }{\partial
\mathbf{q}_{i}},  \label{cr1}
\end{equation}%
or
\begin{equation}
i[m_{i}\frac{\partial }{\partial \mathbf{p}_{i}},\mathcal{L}_{n}]\delta
t=\delta t\frac{\partial }{\partial \mathbf{q}_{i}}.  \label{cr2}
\end{equation}
With definitions of $M$, $\mathbf{R}$ and $\mathbf{p,}$ we obtain
\begin{equation}
i\sum_{i}[m_{i}\frac{\partial }{\partial \mathbf{p}_{i}},\mathcal{L}%
_{n}]=\sum_{i}\frac{\partial }{\partial \mathbf{q}_{i}}  \label{fr}
\end{equation}%
This is a basic result, showing the algebraic condition that has
to be fulfilled by the Liouville operator.

It is worth noting that, taking the Liouville operator to be in
the form $\mathcal{L}_{n}=i\{H_{n},\cdot \}_{n}$ we obtain,
\begin{equation*}
-\sum_{i}(m_{i}\frac{\partial }{\partial \mathbf{P}_{i}}\{H_{n},\cdot
\}_{n}-\{H_{n},\cdot \}_{n}m_{i}\frac{\partial }{\partial \mathbf{p}_{i}}%
)=\sum_{i}\frac{\partial }{\partial \mathbf{q}_{i}}
\end{equation*}%
and hence
\begin{equation*}
-\frac{\partial ^{2}H_{n}}{\partial \mathbf{p}_{i}\partial \mathbf{q}_{i}}%
\frac{\partial }{\partial \mathbf{p}_{i}}+\frac{\partial ^{2}H_{n}}{\partial
\mathbf{p}_{i}^{2}}\frac{\partial }{\partial \mathbf{q}_{i}}=\frac{1}{m_{i}}%
\frac{\partial }{\partial \mathbf{q}_{i}}
\end{equation*}%
or
\begin{equation*}
\frac{\partial ^{2}H_{n}}{\partial \mathbf{p}_{i}\partial \mathbf{q}_{i}}%
=0,\;\;\;\frac{\partial ^{2}H_{n}}{\partial \mathbf{p}_{i}^{2}}=\frac{1}{%
m_{i}},
\end{equation*}%
whose solution is
\begin{equation*}
H_{n}=\sum_{i=1}^{n}\frac{\mathbf{p}_{i}^{2}}{2m_{i}}+V(\mathbf{q}_{1},\cdot
\cdot \cdot ,\mathbf{q}_{n}),
\end{equation*}%
where $V(\mathbf{q}_{1},\cdot \cdot \cdot ,\mathbf{q}_{n})\equiv
V(q)$ is an arbitrary function. Therefore, the condition specified
by Eq.~(\ref{cr2}) gives, in particular, the standard expression
for $\mathcal{L}_{n}$. However, Eq.~(\ref{cr2}) is general and is
satisfied, in principle, for systems where $H$ is not defined.

Using the Galilean invariance, a general equation of motion is
derived
by writing the Lagrangian associated with Eq.~(\ref{liouaug08}%
) in the form
\begin{equation}
\mathfrak{L}=\theta ^{\dagger }\left( i\partial
_{t}+i\frac{p}{m}\partial _{q}+iF(q)\partial _{p}\right) \theta
+g(\theta \theta^{\dagger }), \label{nov181}
\end{equation}%
where $F(q)=-\partial _{q}V(q)$ and $g$ is an arbitrary functional
of the wave functions. This Lagrangian gives rise to
Eq.~(\ref{liouaug08}) with $g=0.$ Let us consider,
as an example, (1+1)-dimensions with $F(q)=0$ and $g(\theta \theta ^{\dagger })=-%
\frac{\lambda }{4}(\theta \theta ^{\dagger })^{2}$, such that
Eq.~(\ref{nov181}) leads to
\begin{equation*}
\left( i\partial _{t}+i\frac{p}{m}\partial _{q}\right) \theta
=\lambda (\theta \theta ^{\dagger })\theta .
\end{equation*}%
This equation describes a flow of particles in phase space,
without external field and with a local non-linear collision term.
Writing $\theta (q,p;t)=\phi
(q,p)\exp (-i\nu t),$ we have%
\begin{equation*}
\left( \nu +i\frac{p}{m}\partial _{q}\right) \phi (q,p)=\lambda \phi
(q,p)^{3}.
\end{equation*}%
The zero-order ($\lambda =0$) solution is $\phi _{0}(q,p)=Ae^{i\nu mq/p}.$
And a solution, up to first order in $\lambda $, is $\phi (q,p)\simeq \phi
_{0}(q,p)+\lambda \phi _{1}(q,p),$ that reads%
\begin{equation*}
\phi (q,p)=\phi _{0}(q,p)+\frac{2A\lambda }{\nu (3i+1)}e^{i3\nu mq/p}.
\end{equation*}%
The distribution function in phase space is
$f(q,p)=\theta^{\dagger }\theta =\phi ^{\dagger }(q,p)\phi (q,p).$
As an illustration, is is important to take the average of an
observable as the momentum, giving rise to the momentum flow.
Following the previous prescription we have,
\[
\langle P\rangle=\int
dqdp\theta^*(q,p;t)\widehat{P}\theta(q,p;t)=\int dqdppf(q,p;t),
\]
that is consistent with the usual result.
\section{Symplectic quantum mechanics}

In this section we consider representations using the
star-product. For simplicity we treat a one-particle system. The
generalization for an $n$ particle system is obtained by following
a procedure
similar to the classical case. The representation space is still $\mathcal{H}%
(\Gamma )$ but equipped with the star-product. The Galilei-Lie
algebra in phase space is constructed by using the operators given
by $f\ast $, according to Eq.~(\ref{A98}). We proceed by selecting
the following set of
functions in $\Gamma $: $p_{i},q_{i},\ell _{i}=\epsilon _{ijk}q_{j}p_{k}$, $%
k_{i}=mq_{i}-tp_{i}$ (sum over repeated indices is assumed). This
set is a
hint to look for Weyl operators fulfilling the Galilei Lie algebra, Eq. (\ref%
{12nov1}); we have,
\begin{eqnarray}
\widehat{P} &=&p*=p -\frac{i\hbar }{2}\partial _{q}\,,
\label{des1} \\
\widehat{Q} &=&q* =q+\frac{i\hbar }{2}\partial _{p},  \label{des2} \\
\widehat{K} &=&k* =mq*-tp*=m\widehat{Q}-t\widehat{P},  \label{des3} \\
\widehat{L}_{i} &=&\epsilon _{ijk}\widehat{Q}_{j}\widehat{P}_{k}  \label{des4} \\
&=&\epsilon _{ijk}q_{j}p_{k}-\frac{i\hbar }{2}\epsilon _{ijk}q_{j}{\frac{%
\partial }{\partial q_{k}}}+\frac{i\hbar }{2}\epsilon _{ijk}p_{k}{\frac{%
\partial }{\partial q_{k}}}+\frac{\hbar ^{2}}{4}\epsilon _{ijk}{\frac{%
\partial ^{2}}{\partial q_{j}\partial p_{k}}},  \label{des5} \\
\widehat{H} &=&i\hbar \frac{\partial }{\partial t}.
\end{eqnarray}
The physical content of this representation is derived, first, by
observing that $\widehat{Q}$ and $\widehat{P}$ are transformed by
the boost according to
\begin{eqnarray}
\exp \left( -iv\cdot \widehat{K}/\mathbb{\hbar }\right) \widehat{Q}_{j}\exp
\left( iv\widehat{K}/\mathbb{\hbar }\right) &=&\widehat{Q}_{j}+v_{j}t\mathbf{%
,}  \label{MAIS509} \\
\exp \left( -iv\cdot \widehat{K}/\mathbb{\hbar }\right) \widehat{P}_{j}\exp
\left( iv\cdot \widehat{K}/\mathbb{\hbar }\right) &=&\widehat{P}_{j}+mv_{j}%
\mathbf{.}  \label{mais6609}
\end{eqnarray}%
Furthermore
\begin{equation}
\left[ \widehat{Q}_{j},\widehat{P}_{l}\right] =i\hbar \delta _{jl}\mathbf{.}
\label{MAIS5-209}
\end{equation}%
Therefore, the operators $\widehat{Q}$ and $\widehat{P}$
correspond to the
physical observables of position and momentum, respectively, with Eqs.~(\ref%
{MAIS509}) and (\ref{mais6609}) describing, consistently, the \ way   $%
\widehat{Q}$ and $\widehat{P}$ transform under the Galilei boost.
The Heisenberg commutation relation is given by
Eq.~(\ref{MAIS5-209}) and $m$
is the mass. As a consequence, and for consistency, the generators $\widehat{%
L}_{i}$ and $\widehat{H}$ are interpreted as the angular momentum
and the Hamiltonian operators, respectively. It is important is to
determine a general form for $H$, that is accomplished by using
the Galilei group. The time evolution of an observable
$\widehat{A}$ is specified by
\begin{equation}
\exp \left( it\widehat{H}/\hbar \right) \widehat{A}(0)\exp \left( -it%
\widehat{H}/\hbar \right) =\widehat{A}(t),  \label{MAIS6}
\end{equation}%
which results in
\begin{equation}
i\hbar {\frac{\partial }{\partial
t}}\widehat{A}(t)=[\widehat{A}(t),\widehat{H}].  \label{MAIS7}
\end{equation}

For a homogeneous system,   the commutation  relations
 $[\widehat{K},\widehat{H}]=i\widehat{P}$, leads to
\begin{equation*}
\lbrack mq+i\frac{\partial }{\partial p},H(q,p)*]=ip+\frac{1}{2}\frac{%
\partial }{\partial q},
\end{equation*}%
where
\begin{equation*}
H(q,p)*=H(q+\frac{i}{2}\frac{\partial }{\partial p}%
,p-\frac{i}{2}\frac{\partial }{\partial q}).
\end{equation*}%
A solution is $H(q,p)=\frac{p^{2}}{2m}+V(q)$. This   result
provides a general functional for the Hamiltonian
\begin{eqnarray*}
H(q,p)* &=&\frac{p*^{2}}{2m}+V(q)* \\
&=&{\frac{p^{2}}{2m}}-{\frac{\hbar ^{2}}{8m}}{\frac{\partial ^{2}}{\partial
q^{2}}}-{\frac{i\hbar p}{2m}}{\frac{\partial }{\partial q}} \\
&&+\widehat{V}(q+{\frac{i\hbar }{2}}{\frac{\partial }{\partial p}).}
\end{eqnarray*}%
It is worthy of noting that this expression for the Hamiltonian
cannot be derived by using
the Casimir invariant of the Galilei-Lie Algebra $I=\widehat{H}-\widehat{P}%
^{2}/2m$. In the next section we set forth a set of rules for  a
complete physical interpretation of the theory in terms of the
notion of states.

\section{Non-linear Schr\"{o}dinger equation in phase space}

Let us introduce a frame in the Hilbert space for the
representation analyzed in the previous section. We define the
operators
\[
\overline{Q}=\widehat{Q}-\frac{i\hbar }{2}\partial _{p} \,\,\,
{\rm{and}}\,\,\, \overline{P}=\widehat{P}+\frac{i\hbar
}{2}\partial _{q}
\]
transform as
\begin{equation}
\exp \left( -\frac{i}{\hbar }vK\right) 2\overline{Q}\exp \left( \frac{i}{%
\hbar }vK\right) =2\overline{Q}+vt  \label{A176-4}
\end{equation}%
and
\begin{equation}
\exp \left( -\frac{i}{\hbar }vK\right) 2\overline{P}\exp \left( \frac{i}{%
\hbar }vK\right) =2\overline{P}+mv.  \label{A176-5}
\end{equation}%
As for the observables $\widehat{P}$ and $\widehat{Q}$ in
Eqs.~(\ref{MAIS5}) and (\ref{mais66}), we find that $\overline{Q}$
and $\overline{P}$ also
transform as position and momentum. However, since $[\overline{Q},\overline{P%
}]=0,\overline{Q}$ and $\overline{P}$ cannot be interpreted as
observables, although they can be used to construct a frame in the
Hilbert space with the content of the phase space. Then we
introduce $\left\vert q,p\right\rangle $  such that
\begin{equation}
\overline{Q}\left\vert q,p\right\rangle =q\left\vert
q,p\right\rangle \,\,\,\, {\rm{and}}\,\,\,\overline{P}\left\vert
q,p\right\rangle =p\left\vert q,p\right\rangle \,,  \label{A180}
\end{equation}%
with
\begin{equation}
\langle q,p\left\vert q^{\prime },p^{\prime }\right\rangle =\delta
(q-q^{\prime })\delta (p-p^{\prime }),  \label{A181}
\end{equation}%
and
\begin{equation}
\int dqdp\left\vert q,p\right\rangle \left\langle q,p\right\vert =1%
 .  \label{A182}
\end{equation}%
Then we have,
\begin{equation}
\psi (q,p,t)=\langle q,p\left\vert \psi ,t\right\rangle .
\label{B1}
\end{equation}%
Here $\psi (q,p,t)$ is a wave function but not with the content of
the usual quantum mechanical state, for $q$ and $p$ are the
eigenvalues of the operators $\overline{Q}$ and $\overline{P}$
which are ancillary variables and not observables.

From Eq.~(\ref{A182}), we have
\begin{equation}
\langle \psi \left\vert \phi \right\rangle =\left\langle \psi
\right\vert \left( \int dqdp\left\vert q,p\right\rangle
\left\langle q,p\right\vert \right) \left\vert \phi \right\rangle
=\int dqdp\psi ^{\dagger }(q,p)\phi (q,p).  \label{B1-2}
\end{equation}%
Using the definition of the star-product, we also have
\begin{equation}
\langle \psi \left\vert \phi \right\rangle =\int dqdp\psi
^{\dagger }(q,p)\ast \phi (q,p).  \notag
\end{equation}

The average of a physical observable $\widehat{A}(q,p)=a(q,p;t)\ast ,$ in
the state $\psi (q,p)$ is given by
\begin{eqnarray}
\langle \widehat{A} \rangle &=&\int dqdp\psi ^{\dagger }(q,p)\widehat{A}%
(q,p)\ \psi (q,p)\   \notag \\
&=&\int dqdp\psi ^{\dagger }(q,p)[a(q,p)\ast \psi (q,p)]  \notag \\
&=&\int dqdp\ a(q,p)[\psi (q,p)\ast \psi ^{\dagger }(q,p)]
\label{bert555}
\end{eqnarray}%
The quantity  $\langle \widehat{A} \rangle $ will be real if the
spectrum of $\widehat{A}$ is real.

The equation of motion is determined by the Lie algebra, resulting
in the Heisenberg-equation in phase space
\begin{equation*}
i\partial _{t}\widehat{A}(q,p;t)=[\widehat{A}(q,p;t),\widehat{H}(\widehat{q},%
\widehat{p})].
\end{equation*}
Therefore,  the Schr\"{o}dinger picture is derived, from the average of $%
\widehat{A}$, that is given by
\begin{eqnarray}
\langle \widehat{A} \rangle &=&{\int }dqdp\,\psi ^{\dagger
}(t)\widehat{A}(0)\psi
(t)  \notag \\
&=&{\int }dqdp\,\psi ^{\dagger }(t)  a(0)\ast \psi (t),
\label{evol44}
\end{eqnarray}%
where $\psi (t)=e^{-i\widehat{H}t}\psi (t).$ Then we obtain the
Shr\"{o}dinger equation in phase space
\begin{eqnarray}
i\hbar \partial _{t}\psi (q,p;t) &=&\widehat{H}(q,p)\psi (q,p;t), \notag  \\
&=&\left( {\frac{p^{2}}{2m}}-{\frac{\hbar ^{2}}{8m}}{\frac{\partial ^{2}}{%
\partial q^{2}}}-{\frac{i\hbar p}{2m}}{\frac{\partial }{\partial q}}\right)
\psi (q,p;t) \\
&&+\widehat{V}\left( q+{\frac{i\hbar }{2}}{\frac{\partial }{\partial p}}%
\right) \psi (q,p;t).
\end{eqnarray}

A fundamental physical result in this formalism is the connection
of $\psi (q,p;t)$ with the Wigner function, $f_{W}(q,p)$, that is
given by
\begin{equation}
f_{W}(q,p)=\psi (q,p)\ast \psi ^{\dagger }(q,p), \label{wig231}
\end{equation}%
fulfilling the Liouville-von Neumann equation~\cite{seb1}. Using
the star-product, the probability density in the configuration
space is defined by
\begin{equation}
\rho (q)=\int dp\,\psi (q,p)\ast \psi ^{\dagger }(q,p)=\int
dp\,\psi (q,p)\psi ^{\dagger }(q,p),  \label{B21}
\end{equation}%
while in momentum space it is
\begin{equation}
\rho (p)=\int dq\,\psi (q,p)\ast \psi ^{\dagger }(q,p)=\int
dq\,\psi (q,p)\psi ^{\dagger }(q,p).  \label{b233}
\end{equation}%
The wave function, $\psi (q,p)$, is then interpreted as a
\textit{quasi-probability amplitude} describing the state of the
system.

The Galilean invariant Lagrangian density for bosons with a
non-linear self interaction is
\begin{eqnarray}
\mathfrak{L} &=&\frac{i\hbar }{2}(\psi ^{\dagger }\partial
_{t}\psi -\psi \partial _{t}\psi ^{\dagger })+\frac{i\hbar
}{4m}p(\psi ^{\dagger }\partial _{q}\psi
-\psi \partial _{q}\psi ^{\dagger })  \notag \\
&&-\frac{p^{2}}{2m}\psi \psi ^{\dagger }+\widehat{V}(q) (\psi \psi ^{\dagger })-%
\frac{\hbar ^{2}}{8m}\partial _{q}\psi \partial _{q}\psi ^{\dagger }+(\psi
\psi ^{\dagger })^{2}.
\end{eqnarray}
Then the Euler-Lagrange equation is%
\begin{eqnarray*}
i\hbar \partial _{t}\psi (t) &=&\left( {\frac{p^{2}}{2m}}-{\frac{\hbar ^{2}}{%
8m}}{\frac{\partial ^{2}}{\partial q^{2}}}-{\frac{i\hbar p}{2m}}{\frac{%
\partial }{\partial q}}\right) \psi (t) \\
&&+\widehat{V}\left( q+{\frac{i\hbar }{2}}{\frac{\partial }{\partial p}}%
\right) \psi (t)+\lambda (\psi ^{\dagger }\psi )\psi .
\end{eqnarray*}%
This describes an extension of the Gross-Pitaevskii equation to
the phase space.

Let us consider, as an example, $\widehat{V}=0$, and $\lambda \ll
1$. Then a linear approximation can be used, i.e. $\psi
(q,p;t)=\psi _{0}(q,p;t)+\lambda \psi _{1}(q,p;t),$ where $\psi
_{0}(q,p;t)$ is the solution \ of the linear equation,
\begin{equation*}
i\hbar \partial _{t}\psi _{0}(t)=\left( {\frac{p^{2}}{2m}}-{\frac{\hbar ^{2}%
}{8m}}{\frac{\partial ^{2}}{\partial q^{2}}}-{\frac{i\hbar p}{2m}}{\frac{%
\partial }{\partial q}}\right) \psi _{0}(t).
\end{equation*}%
For simplicity we analyze the (1+1)-dimensional case. A particular
solution for $\psi _{0}(t)$ is
\begin{equation*}
\psi _{0}(q,p;t)=\phi _{0}(q,p)e^{-iEt/\hbar},
\end{equation*}%
where $\phi _{0}(q,p)=Ae^{k_{\pm }q}$ with%
\begin{equation*}
k_{\pm }=\frac{p}{\hbar }4i[1\pm \frac{1}{p}\left( 2mE\right) ^{1/2}].
\end{equation*}%
In addition
\begin{equation*}
\psi _{1}(q,p;t)=\phi _{1}(q,p)e^{-iEt},
\end{equation*}%
where $\phi _{1}(q,p)=Be^{3k_{\pm }q}$, with
\begin{equation*}
B=\frac{8mA^{3}}{(\hbar k_{\pm })^{2}-i8\hbar k_{\pm }-8mE}.
\end{equation*}
The Wigner function, up to first order in $\lambda ,$ is  given by%
\begin{eqnarray*}
f_{w}(q,p;t) &=&\psi _{0}(q,p;t)\ast \psi _{0}^{\dag }(q,p;t) \\
&&+\lambda \psi _{1}(q,p;t)\ast \psi _{0}^{\dag }(q,p;t) \\
&&+\lambda \psi _{0}(q,p;t)\ast \psi _{1}^{\dagger }(q,p;t).
\end{eqnarray*}%
The star product has to be explicitly developed, providing a
non-trivial result for the Wigner function.

The  average of the momentum, as an example,  is  given by
\[
\langle \widehat P \rangle   = \int
dqdp\,\psi^{\dagger}(q,p;t)\widehat P\psi(q,p;t)=\int dpdq\, p
f_{w}(q,p;t),
\]
 where we have used Eqs.~(\ref{bert555}) and
(\ref{wig231}). Physically, this result is consistent with the
Wigner formalism and describes a quantum flow of bosons in phase
space, with the collision term of the kinetic equation being local
and non-linear.

\section{Conclusion}

In this paper we have studied symplectic (unitary) representations
of the Galilei group for classical and quantum systems, developing
two aspects. First, we have found the general conditions that the
generator of time translation in phase space has to satisfy, such
that its explicit form is derive from general elements of
symmetry. Second, we derive non-linear equations in phase space,
associated with the kinetic theory.

For the classical systems, the generator of time translation is
the  Liouville differential operator, and from the Lagrangian
formalism, a  classical Liouville equation is derived with a local
and non-linear collision term. For quantum systems, the time
generator is a Hamiltonian written in phase space, and the
analysis of the Lagrangian leads to a non-linear Schr\"{o}dinger
equation in phase space. These classical and quantum equations are
solved perturbatively, as an example, to emphasize the usefulness
of such representations in non-relativistic kinetic theory. At the
same time, these results open doors for further developments, such
as the analysis of quantum dynamical systems in phase space.
\\
\textbf{Acknowledgments}:
 The authors thank NSERC (Canada) and CAPES and CNPq (Brazil) for financial support.

\end{document}